\documentclass[preprint,amsmath,amssymb]{revtex4}
\usepackage{graphicx}
\usepackage{dcolumn}
\usepackage{bm}
\begin{document}
\title{General relativistic analogs of Poisson's equation and gravitational binding energy}
\author{${\rm R. \;Gharechahi}^{\;(a)}$ \footnote {Electronic
address:~r.gharechahi@ut.ac.ir}, ${\rm J.\;Koohbor}^{\;(a)}$ \footnote{Electronic
address:~jkoohbor@ut.ac.ir} and ${\rm M.\;Nouri}$-${\rm Zonoz}^{\;(a,b)}$\footnote{Electronic
address:~nouri@ut.ac.ir\; (Corresponding author)}}
\affiliation{(a): Department of Physics, University of Tehran, North Karegar Ave., Tehran 14395-547, Iran.\\ 
(b): School of Astronomy, Institute for Research in Fundamental Sciences, P O Box 19395-5531, Tehran, Iran}
\begin{abstract}
Employing the quasi-Maxwell form of the Einstein field equations in the context of gravitoelectromagnetism, 
we introduce a general relativistic analog of Poisson's equation as a natural outcome of the corresponding spacetime 
decomposition formalism. The active density introduced in this formalism, 
apart from the matter-energy density and pressure, includes a third  component which is the gravitoelectromagnetic 
energy density. This general 
relativistic analog of Poisson's equation is compared with another analog introduced by Ehlers et al. in \cite{Eh1}.
Introduction of the cosmological constant and its effect on the active mass, are also discussed for both exterior and interior static spacetimes.
In the stationary case, we consider the kerr spacetime with a special choice for its interior metric.
\end{abstract}
\maketitle
\section{Introduction and motivation}
One of the main criteria in the development of scientific theories in physics has been the assertion that, any new theory 
should reduce to the old theory it has replaced, in that limit of its parameters where the old theory is expected to be at work. 
In the case of quantum mechanics this is embodied in the celebrated {\it correspondence principle} of Bohr. Einstein used the same 
idea to fix the constant factor appearing in his field equations of  general relativity (GR), namely $ G_{ab} = \kappa T_{ab}$, by looking for a Poisson equation 
in the Newtonian approximation \cite{Steph}. This approximation comprises taking the energy density as the only source of the gravitational 
field in the slowly varying, weak-field limit so that  \footnote{Our convention for indexes is such that Latin indexes run from $0$ to $3$ 
while the Greek ones run from $1$ to $3$. We  use the  $(+,-,-,-)$ signature and set $c=G=1$.} 
\begin{equation}
g_{ab} = \eta_{ab} + f_{ab} 
\end{equation}
with $f_{ab} \ll 1$ and $f_{ab,0} = 0$ leading to 
\begin{equation}
\bigtriangledown^2 f_{00} = \kappa T_{00} \equiv \kappa \rho.  
\end{equation}
This has obviously the form of a Poisson equation, but one should be cautious with the introduction of the correct Newtonian gravitational
potential in this equation. To identify the correct potential one could look at the geodesic equation in the same approximation which leads to
the following equation of motion for a test particle
\begin{equation}
\frac{d^2 x^\mu}{dt^2} \approx 1/2 \eta^{\mu\nu} f_{00,\nu}
\end{equation}
in which $t$ is the coordinate time. Comparing the last equation with the Newtonian equation of motion in a gravitational potential $U$, we end up with 
\begin{equation}
 f_{00} = 2U \;\;\;\;\; ; \;\;\;\;\;  g_{00} = (1+2U).
\end{equation}
The above argument shows the crucial role of the geodesic equation in identifying the correct gravitational potential in the Newtonian limit.\\
Starting from Einstein field equations one can introduce relativistic analogs of Poisson's equation and along with it a gravitational potential. 
Obviously then, there will be  
freedom in choosing a gravitational potential, but this comes at the cost of introducing
new energy-matter content  on the right-hand side of the Poisson equation, with the immediate task of finding their interpretation. \\
Recently a general relativistic analog of Poisson's equation was introduced for {\it static} gravitational fields, where the relativistic potential is defined 
as the potential energy per unit mass \cite{Eh1}. Obviously, in analogy with its  Newtonian counterpart, the right-hand side of such an equation could be identified as 
the source of the gravitational field, which is alternatively called {\it active mass} \cite{Eh1,Eh2}, {\it proper mass} \cite{Wald} or {\it bare mass} \cite{ryder}. 
In their definition, as pointed out by the authors, there is an obvious deficiency in the analogy in that the acceleration on 
a particle at rest in the static gravitational field, is not given by the gradient of the introduced potential. 
Introduction of a relativistic analog of the Poisson equation by its 3-dimensional nature involves a spacetime decomposition formalism, specially 
in the case of stationary spacetimes which possess off-diagonal metric components.
Here employing the so-called {\it gravitoelectromagnetism} (GEM) we introduce a relativistic analog of Poisson's equation for {\it stationary} spacetimes through 
the quasi-Maxwell form of the EFEs. In this definition a third component is added to the active mass density which is the gravitoelectromagnetic energy density. 
We will discuss the two definitions in the context of gravitational binding energy,  which is the difference between the defined
active mass and  the physical mass of a star, as measured, for example, by a planet orbiting the star on a timelike geodesic.
\\The paper is organized as follows. In Sec. II we introduce the $1+3$ or threading formulation of spacetime decomposition and derive the quasi-Maxwell form 
of the Einstein field 
equations in the presence 
of a perfect fluid. In Sec. III  we introduce our general relativistic analog of Poisson's equation in stationary spacetimes and  compare it with 
the one given in \cite{Eh1}. 
In Sec. IV, using the active mass densities in these two definitions, we 
calculate corresponding gravitational active masses in different spacetimes, including  Schwarzschild and de Sitter spacetimes. We also apply our  
formalism to the case of Kerr 
spacetime as the prototype of stationary spacetimes, with a special choice for the interior Kerr solution. Finally, in Sec. V we summarize and 
discuss our main results. 
\section{Gravitoelectromagnetism and the quasi-Maxwell form of the Einstein field equations: A brief introduction}
The $1+3$ or threading formulation of spacetime decomposition is the decomposition of spacetime by the worldlines of {\it fundamental observers} who are at fixed spatial 
points in a gravitational field. In other words, these worldlines sweep the history of their spatial position 
decomposing the underlying spacetime into timelike threads  \cite{Lan}. In stationary asymptotically flat spacetimes, these observers 
are at rest with respect to distant observers
in the asymptotically flat region. Employing radar signal propagation between two nearby fundamental observers (i.e ignoring spacetime curvature),  
the spacetime metric could be expressed in the following general form,
\begin{equation}\label{ds0}
d{s^2} = d\tau_{sy}^2 - d{{l}^2} = {g_{00}}{(d{x^0} - {g_\alpha }d{x^\alpha })^2} - {{\gamma}_{\alpha \beta }}d{x^\alpha }d{x^\beta }
\end{equation}
where ${g_\alpha } =  - \frac{{{g_{0\alpha }}}}{{{g_{00}}}}$ and
\begin{equation}\label{gamma0}
{{\gamma} _{\alpha\beta}} =  - {g_{\alpha \beta }} + \frac{{{g_{0\alpha }}{g_{0\beta }}}}{{{g_{00}}}} \;\; ; \;\;  {{\gamma} ^{\alpha \beta }} =  - {g^{\alpha \beta }}
\end{equation}
is the spatial metric of a 3-space $\Sigma_3$, on which $d{{l}}$ gives the element of spatial distance between any two nearby events.
All the tensor operations in this 3-space are defined with respect to the three-dimensional metric $\gamma_{\alpha\beta}$, and more specifically, the covariant 
differentiation of a 3-vector $T^\alpha$ in this 3-space is defined as follows;
\begin{equation}\label{der}
T^\alpha_{;\beta} = T^\alpha_{,\beta} + \lambda^{\alpha}_{\beta\mu} T^\mu
\end{equation}
in which $\lambda^{\alpha}_{\beta\mu}$ is the Christoffel symbol made out of the  metric $\gamma_{\alpha\beta}$ in the same way that the usual
connection coefficients are made out of the  metric $g_{ab}$.
Also, $d{\tau _{sy}} = \sqrt {{g_{00}}} (d{x^0} - {g_\alpha }d{x^\alpha })$ gives the infinitesimal interval of the so-called {\it synchronized proper time} 
between any two events. In other words any two simultaneous events have a world-time difference of $d x^0 = {g}_\alpha dx^\alpha$. 
In the threading formalism, the 3-velocity for a test particle is measured by {\it fundamental observers} which are at rest with respect to a 
rigid global coordinate system and is defined 
in terms of the synchronized proper time read by clocks synchronized along the particle's trajectory as follows \cite{Lan, MN}
\begin{equation}\label{velo}
{{v}^\alpha } = \frac{{d{x^\alpha }}}{{d{\tau _{sy}}}} = \frac{{d{x^\alpha }}}{{\sqrt {{g_{00}}} (d{x^0} - {g_\alpha }d{x^\alpha })}}
\end{equation}
Obviously, in the case of static spacetimes (i.e., $ g_{0\alpha} = 0$) the above definition 
reduces to the proper velocity defined by $v^\alpha = \frac{1}{\sqrt{g_{00}}}\frac{dx^\alpha} {dx^0}$.\\
Substituting the above definition of 3-velocity in Eq. \eqref{ds0}, one can show the following relation between the proper and synchronized proper times 
\begin{equation}\label{ds00}
d{\tau^2} = {g_{00}}{(dx^0 - g_{\alpha }d{x^\alpha })^2}[1 - {{v}^2}] = d\tau_{sy.}^2 (1-{v}^2)
\end{equation}
Also the components of the 4-velocity $u^i =dx^{i}/d\tau$ of a test particle, in terms of the components of its 3-velocity, are given by
\begin{eqnarray}\label{4u}
u^\alpha = \frac{v^\alpha}{\sqrt{1-v^2}}, \;\; u^0 = \frac{1}{\sqrt{1-v^2}}\left( \frac{1}{\sqrt{g_{00}}}+g_{\alpha}v^{\alpha}\right) 
\end{eqnarray} 
For a test particle moving in a {\it stationary} spacetime, starting from the spatial components of the corresponding  timelike geodesic equation
\begin{eqnarray}\label{geodesic}
\frac{du^\mu}{d\tau} = -\Gamma^\mu_{ab} u^a u^b
\end{eqnarray}
and using  expressions for the connection coefficients in terms of the Three-dimensional objects and  substituting  for the 4-velocity components from \eqref{4u}, 
the left-hand side of the above 
equation could be written as the force acting on the particle defined as the derivative of its momentum with respect to the 
synchronized proper time \cite{Lan,MN},
\begin{equation}\label{force1}
{f}^\mu \equiv  \frac{D P^\mu}{d\tau_{sy}} = {\sqrt{1-v^2}} \frac{d}{d\tau}\frac{mv^\mu}{\sqrt{1-v^2}} + \lambda^{\mu}_{\alpha\beta} \frac{mv^\alpha v^\beta}{\sqrt{1-v^2}}
\end{equation}
Intuitively, this shows that test particles
moving on the geodesics of a stationary spacetime depart from the geodesics of the 3-space
$\Sigma_3$ as if acted on by the above GEM Lorentz-type 3-force.
In its vectorial form (with lowered index), it could be written after a long but straightforward manipulation as follows
\begin{equation}\label{force}
{\bf f}_g =  \frac{m_0}{\sqrt{1-{v^2}}}\left( {\bf E}_g + {\bf v}\times \sqrt{g_{00}}{\bf B}_g\right)
\end{equation}
in which the gravitoelectric (GE) and gravitomagnetic (GM) 3-fields are defined as follows \footnote{It should be noted that all the differential
operations in these relations are defined in the 3-space $\Sigma_3$ with metric $\gamma_{\mu\nu}$.};
\begin{gather}
\textbf{B}_g = curl~({\bf A}_g)\;\;  ; \;\; ({A_g}_\alpha \equiv g_{\alpha})\label{bg} \\
\textbf{E}_g = -{\bf {\nabla}} \ln \sqrt{h} \;\;  ; \;\; (h \equiv  g_{00})\label{pot} .
\end{gather}
Obviously they satisfy the following constraints
\begin{equation}
\nabla \times~\textbf{E}_g=0, ~~~\nabla \cdot  \textbf{B}_g=0. 
\end{equation}
Now in terms of the GEM fields measured by the fundamental observers 
\footnote{In the case of stationary spacetimes these observers move along the timelike geodesics and are called Killing observers.},
Einstein field equations for a one-element perfect fluid could be written 
in the following quasi-Maxwell form \cite{MN},
\begin{gather} 
\nabla \cdot \textbf{E}_g= \frac{1}{2} h B^2_g+E^2_g - {8\pi}\left(\dfrac{p+\rho}{1-v^2}-\dfrac{\rho-p}{2}\right) \label{r00} \\  
\nabla \times  (\sqrt{h}\textbf{B}_g)=2 \textbf{E}_g \times (\sqrt{h}\textbf{B}_g)-{16\pi}\left(\dfrac{p+\rho}{1-v^2}\right) {\textbf{v}} \label{r01} \\
{^{(3)}}P^{\mu\nu}=-{E}_g^{\mu;\nu}+\frac{1}{2}h(B_g^\mu B_g^\nu - B_g^2 \gamma^{\mu\nu})+ {E}_g^\mu E_g^\nu+ 
{8\pi}\left(\dfrac{p+\rho}{1 - v^2}v^\mu v^\nu+\dfrac{\rho-p}{2}\gamma^{\mu\nu}\right) \label{3ricci}
\end{gather}
in which $\bf v$ is the 3-velocity of the perfect fluid as defined in (\ref{velo}) and ${^{(3)}}P^{\mu\nu} $ is the three-dimensional Ricci 
tensor of the 3-space $\Sigma_3$.\\
The above formalism can be employed to derive gravitational analogs of some of the electromagnetic effects such as those studied in \cite{MN, Nouri99, Nouriz, Fil}. 
Here, what concerns us mainly is the following point:
Comparing Eq. \eqref{r00} with its analog in electromagnetism, one could define the $GEM$ energy density measured by fundamental observers in the following form
\begin{equation}
u_g = \frac{1}{2} h B_g^2 + E_g^2 \label{energy}
\end{equation}
In what follows, we will discuss the consequences of the presence of the above quantity in the general relativistic analog of Poisson's equation 
introduced in the next section.
\section{General Relativistic analogs of Poisson's equation from Gravitoelectromagnetism}
The Poisson equation in Newtonian theory of gravity $(\nabla^{2}\varphi=4\pi G \rho)$, which is formally the same as its counterpart in electromagnetism, gives the 
relation between 
the gravitational potential and  density of the {\it active mass} which produces the corresponding gravitational field in an inertial coordinate system.  
Looking for a general relativistic analog of Poisson's equation, the above GEM formulation of spacetime decomposition introduces a natural candidate. 
From Eqs. (\ref{force}) and (\ref{pot}),
for a test particle at rest (i.e., in the comoving frame), the force acting on the mass introduces the following definition of GE potential,
\begin{equation}
{\varphi}_{GEM} = \ln\sqrt{h}\label{potential}.
\end{equation}
The same definition could also be obtained solely from the analogy between the definitions of an electric field in terms of a scalar potential and 
its gravitational analog, namely, (\ref{pot}). Looking for the static, spherically symmetric, interior  solutions of Einstein's field equations with a perfect 
fluid, Wald also takes the above form as the general relativistic analog of Newtonian potential \cite{Wald}.  \\
Employing  the above definition of GE potential and using \eqref {energy}, one could express the relation (\ref{r00}) in comoving coordinate system to end up with,
\begin{equation}
{{\nabla}^2} {\varphi}_{GEM}={4\pi} \left( \rho+3p\right) -u_g \label{poi1}
\end{equation} 
The above equation  could be taken as a general relativistic analog of Poisson's equation in {\it stationary spacetimes}. Comparison with the Poisson 
equation in Newtonian gravity shows 
that the expression $\rho_{GEM}=\rho+3p-u_{g}/4\pi$ could be considered as the active mass density of the gravitational field. Obviously, by the general 
form of GEM energy density $u_g$, its contribution to the active mass is expected to be negative.\\
Going back to our starting point, i.e  equation \eqref {geodesic}, the above definition was built to be consistent with the geodesic equation. 
But obviously it is not the only general relativistic analog to Poisson's equation. 
In \cite{Eh1}, following the Newtonian case and restricting attention to {\it static} gravitational fields, Ehlers et.al define 
the general relativistic potential as the work required
to displace a unit mass from infinity. Based on this definition of potential, the authors introduce their general relativistic analog of Poisson's equation 
which we call the EOS definition/formulation.\\
Here we use the threading formulation of spacetime decomposition introduced in the previous section, to show that 
there is no need to restrict
the discussion to static spacetimes and the same definition could be obtained for stationary spacetimes as a particle's potential energy per unit mass in its 
comoving frame \cite{Lan}.
To do so, using definitions of the 3-velocity \eqref {velo}
and 4-velocity \eqref{4u}, one could show 
that in a {\it stationary} spacetime, the energy of a particle defined as the time component of its 4-momentum is given by,
\begin{eqnarray}\label{appa62}
\varepsilon \equiv P_0 =   m g_{0i} u^i = \frac{m \sqrt{g_{00}}}{\sqrt{1-v^2}},\nonumber
\end{eqnarray}
which is a conserved quantity  during the motion of the particle \footnote{This could also be obtained by noting that in a stationary spacetime with a 
timelike Killing vector field $\xi_i$, the quantity $\xi_i u^i$ is a conserved quantity representing the energy of a unit-mass particle.}. In the comoving 
frame it reduces to $m \sqrt{g_{00}}$ \cite{Lan}, leading to the following 
definition of a general 
relativistic gravitational potential 
\begin{equation}
{\varphi}_{EOS} \equiv \frac{\varepsilon (v=0)}{m} = \sqrt{g_{00}} \equiv \sqrt{h}\label{pot2}
\end{equation}
Using the above definition of gravitational potential, one can rewrite Eq. (\ref{r00}) for {\it static spacetimes}
(i.e., ${\bf B}_g=0$) in the comoving coordinates to arrive at 
the following analog of Poisson's equation \footnote{It is interesting to compare this simple physical derivation  with the one which was obtained in a much 
more complicated way in \cite{Eh1}.},
\begin{equation}
{{\nabla}^2}{\varphi}_{EOS}={4\pi}\sqrt{h}\left( \rho+3p\right) \label{poi2}
\end{equation} 

Comparing the above two expressions as the general relativistic analogs of Poisson's equations, the following points are worthy to be mentioned:\\
(1) Since we are using nothing but the Einstein field equations to arrive at either of the above analogs of Poisson's equations \eqref{poi1} and \eqref {poi2}, 
it is obvious that they both reduce to the Newtonian case in the corresponding limit.\\
(2) Active mass densities are given by the expressions ${\rho_{GEM}}=\rho + 3p-u_{g}/4\pi$ and ${\rho_{EOS}} = \sqrt{h}\left( \rho+3p\right)$, respectively. \\
(3) Using relations (\ref{force}) and (\ref{pot}), in the comoving frame, the acceleration of a test particle is given 
by ${a}^{GEM}_\mu \equiv {E_g}_\mu = -\partial_\mu \ln \sqrt{h}\equiv-\partial_\mu {\varphi}_{GEM}$,
whereas taking (\ref{pot2}) as the gravitational potential, the acceleration of a particle at rest 
is ${a}^{EOS}_\mu = -\frac{1}{\sqrt{h}} \partial_\mu \sqrt{h} \equiv - \frac{1}{\sqrt{h}} \partial_\mu {\varphi}_{EOS}$;
i.e., it is not given by the gradient of the potential \cite{Eh1}. \\
(4) For a static spherical source in the weak-field limit, where $g_{00}\approx (1 + {2\Phi_{N}}) = (1 - {2M/r})$, at large distances from the 
center of the source the two definitions of the general relativistic potentials (to the first order in $\Phi_{N}$) are given by; 
\begin{equation}
{\varphi}_{GEM}  = \ln \sqrt{h} \approx - \frac{M}{r} = \Phi_{N} \label{pot12}
\end{equation}
\begin{equation}
{\varphi}_{EOS} = \sqrt{h} \approx 1 - \frac{M}{r} = 1 + \Phi_{N} \label{pot22}
\end{equation}
Obviously, the EOS potential asymptotically approaches the rest mass energy  at the spatial infinity \cite{Eh1}, whereas the GEM potential is the same as 
the Newtonian potential and vanishes asymptotically.
\subsection{Inclusion of the cosmological constant term} 
In the presence of the cosmological constant term, generalization of the above general relativistic Poisson's equation  could be obtained by considering 
the cosmological $\Lambda$-term as the energy-momentum tensor of 
a (dark) perfect fluid with the equation of state $\rho=-p=\Lambda/8\pi$. So we only need to replace the term $\rho+3p$ with $\rho+3p - \dfrac{\Lambda}{4\pi}$ in 
Eqs. \eqref {poi1} and \eqref {poi2} leading to
\begin{equation}
{{\nabla}^2} \varphi_{GEM} = {4\pi} \left(\rho+3p - \dfrac{u_g}{4\pi} - \dfrac{\Lambda}{4\pi}\right) \label{poi4}
\end{equation} 
and 
\begin{equation}
{{\nabla}^2}  \varphi_{EOS} = {4\pi}\sqrt{h}\left( \rho+3p-\dfrac{\Lambda}{4\pi}\right)  \label{poi3}
\end{equation} 
respectively. Compared with Poisson's equation, on the right-hand side of the above equations, both  pressure and the cosmological $\Lambda$-term 
contribute to the active mass. 
Also, in the GEM formulation there appears the GEM energy density as another element contributing to the active mass. In the EOS formulation the 
factor $\sqrt{h}$  on the right-hand side is another obvious difference compared to Poisson's equation. \\
Before advancing further to compare the above two definitions in the next section, which is the main objective of our study, it should be clarified that 
the main concern of the authors in \cite{Eh1} and \cite{Eh2} 
was to revisit the so-called Tolman's paradox \cite{Tolman}. This paradox arises from considering the consequences of 
the pressure term in \eqref{poi2} as a source for the gravitational potential and could be illustrated by taking a static spherical source of matter which could 
undergo an internal transformation. If this transformation is accompanied by a change of the equation of state, such as in the case of matter-antimatter annihilation,
that would obviously result in a
change in the active mass as the source of the gravitational potential. On the other hand, such a change should not affect 
the mass measurement by gravitational means such as light bending or 
massive particle orbits around such a source due to the spherical symmetry and Birkhoff's theorem. The original resolution to the paradox was given by Misner 
and Putnam \cite{Misner} who showed, neglecting gravitational effects, that any change in the 
pressure inside the source is compensated by the stress changes on the boundary of the source, so that their corresponding contributions to the active mass are canceled out. 
Generalizing the result of Misner and Putnam, the authors in \cite{Eh2} consider the same resolution in the case in which gravitational effects are taken 
fully into account. Also, it is noted that as long as Tolman's paradox is concerned, starting from stationary Einstein field equations for perfect fluids, 
the $3p$ term will be present in any general relativistic analog of Poisson's equation, including \eqref{poi1}, and hence, the same resolution would 
be effective for {\it static} spacetimes.   
\section{Active mass and gravitational binding energy from General relativistic analogs of Poisson's equation} 
In this section, to better compare the above two relativistic analogs of the Poisson equation,  we apply them to calculate the total 
active mass in different static and stationary spacetimes.    
\subsection{Schwarzschild spacetime}
The simplest interior solution of the Einstein field equations for a static and spherically symmetric distribution  of matter (say, a star) is the 
interior Schwarzschild solution 
which is obtained for a uniform density $\rho=const$. 
The metric of the interior Schwarzschild solution for a spherical distribution with radius $R$ for $r \leq R$ in Schwarzschild-type coordinates is given by \cite{ryder}
\begin{equation}
ds^2={\dfrac{1}{4}}(3a_0-a)^2dt^2-\dfrac{dr^2}{a^2}-r^2d\Omega^2,  \;\;
a^2=1-\dfrac{8\pi}{3}\rho r^2 \label{isch}
\end{equation}
where $a_0=a(R)$. If $\rho=const$, from the $TOV$ equation the pressure at a radius $r$ is found to be \cite{ryder}
\begin{equation}
p(r)=\dfrac{\rho(a-a_0)}{3a_0-a} \label{Pes}
\end{equation}
which, as expected, vanishes at the surface of the star at $r=R$. On the other hand, from Birkhoff's theorem the spacetime geometry outside a  spherically 
symmetric matter distribution is necessarily static and is given by the Schwarzschild line element
\begin{equation}
ds^2=(1-\frac{2M}{r})dt^2-(1-\frac{2M}{r})^{-1}{dr^2}-r^2 d\Omega^2
\end{equation}
From matching the two metric components at the star's surface $(r=R)$, we immediately find that
\begin{equation}
M = \dfrac{4\pi}{3} R^3 \rho . \label{mass1}
\end{equation}
In other words, the mass parameter in the exterior Schwarzschild solution is equal to the mass contained within a coordinate radius $R$ in Newtonian gravity. 
However one should be careful with  
this analogy in a curved background with  metric (\ref{isch}), where the volume element of a spherical shell of thickness $dr$  
is given by $dV=\sqrt{-g_{rr}}\; 4\pi r^2 dr$. So, within a coordinate radius ${\bar r}$, the active mass producing the gravitational field is given by
\begin{equation}
{\cal M} = 4\pi \int^{\bar r}_{0} {\sqrt{-g_{rr}} \; \rho_{act} r^2  dr} \label{int01}
\end{equation}
Now looking for the active mass of a spherical distribution, producing Schwarzschild geometry, in the EOS formulation, we substitute for the active mass density 
from (\ref{poi2}) to arrive at
\begin{equation}
{{\cal M}}_{EOS} = 4\pi \int^{R}_{0} {\sqrt{g_{00}}(\rho+3p) \sqrt{-g_{rr}}r^2 dr} = \dfrac{4\pi}{3} R^3 \rho \equiv M \label{int1}
\end{equation}
in which from  Eqs. (\ref{isch}) and (\ref{Pes}) we used the fact that for the interior Schwarzschild solution $\sqrt{g_{00}}(\rho+3p)=\rho a$ and $\sqrt{-g_{rr}}=1/a$. 
In other words, in this case the EOS active mass is equivalent to the mass parameter of the external metric, which, in turn, is given by 
the Euclidean volume and the Newtonian matter density.\\
Now we  calculate the gravitational active mass in the GEM formalism based on (\ref{poi1}), by  substituting for the
gravitoelectric fields $\textbf{E}_g$ of the interior and exterior Schwarzschild geometries [using definition (\ref{pot})], to  
find the corresponding density as follows:
\begin{equation}
\rho_{GEM} \equiv \rho + 3p - E_g^2/4\pi= \begin{cases}
 \; \dfrac{2a\rho}{3a_0-a}-\dfrac{M^2 r^2}{\pi R^6(3a_0-a)^2} \;\;\;\;\;\;\; r<R \\ 
 \; -\dfrac{M^2}{4\pi r^4}\left(1- \dfrac{2M}{r}\right)^{-1}  \;\;\;\;\;\;\;\;\;\;\;\;\; r>R
\end{cases}\nonumber 
\end{equation}
so that the gravitational active mass for the interior and exterior solutions are found to be
\begin{gather}
{{\cal M}}_{GEM}^{int} = \int^{R}_{0} \dfrac{\rho_{GEM (r<R)}}{a} 4\pi r^2dr = M(1-\frac{2M}{R})^{-\frac{1}{2}} \approx M + \frac{M^2}{R} + 
\frac{3}{2} \frac{M^3}{R^2} +...,\label{intSc} \\
{{\cal M}}_{GEM}^{ext} =\int^{\infty}_{R} \dfrac{\rho_{GEM (r>R)}}{(1-2M/r)^{1/2}} 4\pi r^2dr = M - M(1-\frac{2M}{R})^{-\frac{1}{2}} 
\end{gather}
in which the result of the first integral is expanded in terms of $M/R \ll 1$ \footnote{This is true for 
typical stars where the star radius
is much larger than its Schwarzschild radius, e.g for sun this ratio is $2\times 10^{-6}$.} to enable a comparison with the results of other definitions of active mass. 
The (total) GEM active mass obviously sums up to
\begin{equation}
{{\cal M}}_{GEM}= {{\cal M}}_{GEM}^{int} + {{\cal M}}_{GEM}^{ext} = M 
\end{equation}
which shows that  the total GEM  active mass is equal to EOS active mass, which was found to be the mass parameter in the exterior Schwarzschild metric. 
On the other hand, if we only focus on the interior solution and the corresponding active mass, then
the extra terms starting with  $\frac{M^2}{R}$ could be taken as the (interior) {\it  gravitational binding energy}. To the leading order, this is larger 
than $\frac{3}{5} \frac{M^2}{R}$, which is the gravitational 
binding energy calculated in the Newtonian approximation of the interior Schwarzschild spacetime, with the {\it constant} matter density taken as the only 
source of active mass \cite{ryder}. \\ 
Obviously, the above considerations show that the EOS formulation of active mass does not accommodate such a thing as gravitational binding energy.
\subsection{de Sitter spacetime} 
To examine the above two different relativistic analogs of Poisson's equation and their corresponding formulas for the active mass, we now turn to the cosmological case 
and the important question of the energy content of the Universe and its effect on the dynamics of the Universe. To do so, we start with the simplest model, the de Sitter 
solution, which by recent observations in the context of the $\Lambda$CDM model seems to be the late time geometry of our Universe. In its static form (i.e., in a 
noncomoving frame), it is given by
\begin{equation}
ds^2 = (1- \dfrac{\Lambda}{3} r^2) dt^2 - (1- \dfrac{\Lambda}{3} r^2)^{-1} dr^2  - r^2d\Omega^2 \label{des}
\end{equation}
One can think of the  de Sitter metric as the solution to the Einstein field equations in the presence of a perfect fluid  with equation of state $\rho=-p=\Lambda/8\pi$.
Indeed, it was shown that the de Sitter solution is the only spherically symmetric, static one-element perfect fluid solution of EFEs in a noncomoving frame \cite{NKR}.
Again, one could use either of the potentials and the corresponding gravitational active mass densities introduced in  \eqref{poi4} or \eqref{poi3} (of course, only 
in the presence of $\Lambda$). Employing the EOS formulation from \eqref{poi3}, we have $\rho_{EOS} = \sqrt{h}(-\Lambda/4\pi)$ so that using \eqref{int01}
for a sphere with radius $R$, we end up with
\begin{equation}
{\cal M}_{EOS} = 4\pi \int^{R}_{0} \sqrt{g_{rr}} \sqrt{h}(-\frac{\Lambda}{4\pi}) r^2  dr = -\dfrac{1}{3}\Lambda R^3 \label{mass5}
\end{equation}
whereas in the GEM formalism with potential  \eqref {poi4}, the corresponding active mass density 
\begin{equation}
\rho_{GEM}=- \dfrac{\Lambda}{4\pi}-\dfrac{1}{4\pi}\dfrac{H^4r^2}{1-H^2r^2}\;, \;\;\;\;\;\;H^2=\dfrac{\Lambda}{3}\nonumber
\end{equation}
leads,  by integration up to a sphere of radius $R$, to the following active mass contained in that sphere
\begin{equation}
{\cal M}_{GEM} = \int^{R}_{0} \sqrt{g_{rr}} \rho_{GEM} 4\pi r^2dr =- \dfrac{H^2R^3}{\sqrt{1-H^2R^2}}\simeq -\dfrac{1}{3}\Lambda R^3 -\dfrac{1}{2}H^4 R^5 \nonumber
\end{equation}
where we have employed expansion with respect to $HR \ll 1$. To the first order, the two results agree  and
the difference between the two active masses could be traced back to the presence of the GEM energy density on the right-hand side of \eqref{poi4}.
Obviously, the de Sitter space is devoid of any matter in the usual sense, and active mass here only accounts for the spacetime's  gravitational
field energy assigned to $\Lambda$ as a {\it dark fluid} \cite{NKR}. On the other hand, since there is no notion of assembly in the case of repulsive gravity represented by 
the cosmological constant,  calling  the second term 
on the right-hand side of the above equation {\it gravitational disperssing energy} seems to be more appropriate. 
\subsection{Schwarzschild-de Sitter spacetime} 
To account for both notions of assembly and dispersion, in this section we consider the contributions of the energy density and the cosmological constant to the active mass density 
in the context of interior and exterior Schwarzschild-de Sitter (SdS) solutions.
Following the same scenario applied to the de Sitter solution in the previous section, the interior solution of Einstein's equations in the presence of the cosmological 
constant for a static and spherically symmetric configuration of uniform density, can be
obtained by substituting  $\rho_{eff}=\rho + \Lambda/8\pi$ and $p_{eff}=p-\Lambda/8\pi$ for $\rho$ and $p$, respectively, in a typical internal Schwarzschild solution.
In this way the line element of an interior SdS spacetime is given by \cite{stu}
\begin{gather}
ds^2= \left[\dfrac{3\rho a_0-(\rho-\Lambda/4\pi)a}{2(\rho+\Lambda/8\pi)}\right] ^2dt^2-\dfrac{dr^2}{a^2}-r^2d\Omega^2 \;\; , 
\;\; a^2=1-\dfrac{1}{3}(8\pi\rho+\Lambda)r^2 \label{int2}
\end{gather} 
Similarly, from the modified $TOV$ equation and the assumption that at the surface of the sphere, the pressure $p$ is zero [i.e., $p(R)=0$], 
the pressure at a radius $r$ is given by the following relation \cite{stu}
\begin{equation}
p=\dfrac{\rho(\rho-\Lambda/4\pi)(a-a_0)}{3\rho a_0-(\rho-\Lambda/4\pi)a} \label{p2}
\end{equation} 
As in the Schwarzschild case, matching the above interior SdS solution with the exterior SdS 
\begin{equation}\label{sch-ds}
ds^2=\left(1-\dfrac{2M}{r}-\dfrac{\Lambda}{3}r^2\right)dt^2-\left(1-\dfrac{2M}{r}-\dfrac{\Lambda}{3}r^2\right)^{-1}dr^2-r^2d\Omega^2
\end{equation}
at the surface of the configuration (i.e., $r=R$), we end up with a similar relation as in \eqref {mass1}.
Again, one could employ the relations \eqref{poi4} and \eqref{poi3} to calculate the active mass generating the gravitational field in SdS spacetime in both formalisms.
Starting with the EOS formalism and the active mass density \eqref{poi3} and substituting for the pressure inside the configuration from
\eqref{p2}, we arrive at the following active mass up to a coordinate radius ${\bar r} > R$, 
\begin{gather}
{\cal M}_{EOS} = 4\pi \int^{\bar r}_{0}{\sqrt{g_{00}}(\rho + 3p - \Lambda/4\pi ) \sqrt{-g_{rr}}{r}^2 dr} \nonumber
\\=4\pi \int^{R}_{0}(\rho - \Lambda/4\pi ){r}^2 dr+4\pi \int^{\bar r}_{R}(- \Lambda/4\pi ){r}^2 dr = M-\dfrac{1}{3}\Lambda {\bar r}^3. \label{mass4}
\end{gather} 
Noting that here  $M$  is the mass parameter in the exterior Schwarzschild-de Sitter solution, as expected, up to the coordinate radius of the star (i.e at $r=R$), the 
active mass is the summation of the results \eqref {int1} and \eqref {mass5}. \\
To calculate the GEM active mass, we start with the GEM active mass density \eqref{poi4} which in this case takes the following form
\begin{equation}
\rho_{GEM} =\begin{cases}
 \; \dfrac{2\left(\rho-\dfrac{\Lambda}{4\pi}\right)\left(\rho+\dfrac{\Lambda}{8\pi}\right)a}{3\rho a_{0}-\left(\rho-\dfrac{\Lambda}{8\pi}\right)a}-
 \dfrac{1}{4\pi}\left[ \dfrac{\left(\rho-\dfrac{\Lambda}{4\pi}\right) \dfrac{8\pi}{3} \left(\rho+
 \dfrac{\Lambda}{8\pi}\right)  r}{3\rho a_{0}-\left(\rho-\dfrac{\Lambda}{8\pi}\right)a} \right]^{2}  , \;\;\;\;\;\;\; r<R\\ 
 \; -\dfrac{1}{4\pi}\left( \dfrac{M}{r^2}-\dfrac{\Lambda}{3}r\right)^{2} \left(1- \dfrac{2M}{r}-\dfrac{1}{3}\Lambda r^2\right)^{-1} 
 ,\;\;\;\;\;\;\;\;\;\;\;\;\;\;\;\;\;\;\;\;\;\;\;\;\;\;\;\;\;\;\;R<r<R_H
\end{cases} 
\end{equation}
in which $R_H = {\dfrac{2}{\sqrt{\Lambda}}}\cos\left[{\dfrac{1}{3}}\cos^{-1}(3M\sqrt{\Lambda})+\dfrac{\pi}{3}\right]$ is the de Sitter horizon. 
The gravitational active mass
up to a coordinate radius $\bar r > R$ is given by
\begin{equation}
{\cal M}_{GEM} = \int^{R}_{0} \rho_{GEM (r<R)} dV + \int^{\overline{r}}_{R} \rho_{GEM (r>R)}dV \;\;\; ; \;\;\; dV=4\pi r^{2}dr \sqrt{-g_{rr}}
\end{equation}
At coordinate radii $\bar r$ where both  $\dfrac{M}{\bar r} \ll 1\; {\rm and} \; \dfrac{\bar r}{R_H} \approx {\bar r} {\sqrt \Lambda} \ll 1 $ are satisfied, 
the above integrals could be calculated to yield,
\begin{equation}
{\cal M}_{GEM} = M + \frac{M^2}{\bar r} - \dfrac{1}{3}\Lambda R^{3}-\dfrac{8}{15}M \Lambda R^{2} -\dfrac{113}{70}\Lambda R  M^2  + 
\dfrac{1}{3}M \Lambda {\overline{r}}^{2} + \dfrac{3}{2}  \Lambda {\overline{r}} M^2 +  {\rm higher \; orders}
\end{equation}
It is noted that terms including mass alone source the attractive gravity but those including both the cosmological constant and mass could account for both 
gravitational binding and dispersing energies.
\title{Active mass of the kerr spacetime}
\section{Active mass of a rotating star}
To examine the active mass in physically interesting stationary spacetimes we do not have that much choice, and obviously the main example should include a rotating 
source with the external geometry described  by the Kerr metric. To distinguish between any two different active mass definitions, we need to match this metric
with an interior Kerr metric at the surface of the source. There are a few  analytic interior Kerr solutions, none of which reduce to the well-known 
interior Schwarzschild solutions when the angular momentum parameter is set equal to zero. To this end, we choose an analytic interior Kerr metric introduced by 
G\"{u}rses and G\"{u}rsey in \cite{Gi}, which matches the Kerr metric on the surface of
an oblate spheroidal source. In Boyer-Lindquist coordinates, this solution is given by the following line element,
\begin{equation}\label{11}
d{s^2} = (\frac{{ {\Delta  - {a^2}{{\sin }^2}\theta } }}{{{\rho ^2}}})d{t^2} + \frac{{4af{{\sin }^2}\theta }}{\rho ^2}dtd\phi 
- \frac{{{\rho ^2}}}{\Delta}d{r^2} - {\rho ^2}d{\theta ^2} - \frac{{B{{\sin }^2}\theta }}{{{\rho ^2}}}d{\phi ^2}
\end{equation}
in which
\begin{equation}
\begin{array}{*{20}{c}}
{{\rho ^2} = {r^2} + {a^2}{{\cos }^2}\theta }\\
{\Delta  = {r^2} + {a^2} - 2f(r)}\\
{\Sigma = {{({r^2} + {a^2})}^2} - {a^2}\Delta {{\sin }^2}\theta }
\end{array}
\end{equation}
where $a$ is the angular momentum parameter (per unit mass)
introduced in the Kerr metric. This metric matches the Kerr metric at the hypersurface $r=R$, denoting the spheroid $\frac{x^2 + y^2}{R^2+a^2} + \frac{z^2}{R^2} =1 $  
only with the function $f(r)$ satisfying the boundary conditions
\begin{equation}\label{BC}
\begin{array}{*{20}{c}}
{f({R}) = M{R}}\;\;\;\; ; \;\;\;\;{f'({R}) = M}
\end{array}
\end{equation}
where the prime denotes differentiation with respect to $r$, and $M$ is the (exterior) Kerr mass parameter.
From Einstein field equations the corresponding stress-energy tensor is given by
\begin{equation}\label{EMT}
{T_{\mu \nu }} = (D + 4H){u_\mu }{u_\nu } - (D + 4H)({{{\rho ^2}} \mathord{\left/
 {\vphantom {{{\rho ^2}} \Delta }} \right.
 \kern-\nulldelimiterspace} \Delta }){m_\mu }{m_\nu } - (D + 2H){g_{\mu \nu }},
\end{equation}
where
\begin{equation}\label{BC1}
\begin{array}{*{20}{c}}
{{u_\mu } = \sqrt {{\Delta  \mathord{\left/
 {\vphantom {\Delta  {{\rho ^2}}}} \right.
 \kern-\nulldelimiterspace} {{\rho ^2}}}} (1,0,0, - a{{\sin }^2}\theta )}\\
{{m_\mu } = (0, - 1,0,0)}\\
{D =  - {{f''} \mathord{\left/
 {\vphantom {{f''} {8\pi {\rho ^2}}}} \right.
 \kern-\nulldelimiterspace} {8\pi {\rho ^2}}}},\\
{{{H = \left( {rf' - f} \right)} \mathord{\left/
 {\vphantom {{H = \left( {rf' - f} \right)} {8\pi {\rho ^4}}}} \right.
 \kern-\nulldelimiterspace} {8\pi {\rho ^4}}}}.
\end{array}
\end{equation}
Now that we have a more general energy-momentum tensor in the Boyer-Lindquist coordinate system, to calculate the gravitational active mass 
in the GEM formulation based on the active density,
we need to go back to the original Einstein field equations
that led to \eqref{r00}, namely $R_{00} = 8\pi ( T_{00} - \frac{1}{2} g_{00} T)$. Using this form to extract the energy-matter content 
of the  GEM active mass density we end up with
\begin{equation}
{\rho _{GEM}}(r < {R}) = ( \frac{2 T_{00}}{g_{00}} - T) - \frac{1}{{4\pi }}{u_g},
\end{equation}
from which by substituting for $T_{00}$ and $T$ from the energy-momentum tensor \eqref{EMT}, we are led to the following active mass density in the GEM formalism,
\begin{equation}\label{rho}
{\rho _{GEM}} = D + \frac{{2\Delta (D + 4H)}}{{\left( {\Delta - {a^2}{{\sin }^2}\theta } \right)}} - (D + 4H)\frac{{\Delta B - \Delta {a^2}{{\sin }^2}\theta 
\left( {\Delta - {a^2}{{\sin }^2}\theta } \right)}}{{B\left( {\Delta  - {a^2}{{\sin }^2}\theta } \right)}} - \frac{1}{{4\pi }}\left( {\frac{1}{2}{g_{00}}{B_g}^2 + {E_g}^2} \right).
\end{equation}
Using the definitions of the  gravitoelectromagnetic fields \eqref{bg} and \eqref{pot} for the Kerr metric 
and after a lengthy but straightforward calculation, we end up with the following results for the
squares of the GE and GM fields,
\begin{equation}
\begin{array}{l}
{E_g}^2 \equiv \gamma^{\alpha\beta} E_{g_{\alpha}} E_{g_{\beta}} = \left. {\frac{{{a^4}f{{(r)}^2}{{\sin }^2}\theta }}{{\left( {{r^2} + {a^2}{{\cos }^2}\theta } \right)
{{\left( {{a^2} + {r^2} - 2f(r) - {a^2}{{\sin }^2}\theta } \right)}^2}}}} \right. + \\
\frac{{\left( {4rf(r) - {{\left( {{a^2} + 2{r^2} + {a^2}\cos 2\theta f'(r)} \right)}^2}} \right)}}
{{4\left( {{r^2} + {a^2}{{\cos }^2}\theta } \right)\left( {{a^2} + {r^2} - 2f(r)} \right){{\left( {{a^2} + {r^2} - 2f(r) 
- {a^2}{{\sin }^2}\theta } \right)}^2}}},
\end{array}
\end{equation}
and
\begin{equation}
\begin{array}{l}
{B_g}^2 \equiv \gamma^{\alpha\beta} B_{g_{\alpha}} B_{g_{\beta}} = \frac{{2{a^2}\left( {{r^2} + {a^2}{{\cos }^2}\theta } \right){{\csc }^2}\theta \left( {{a^2} + {r^2} - 2f(r) - 
{a^2}{{\sin }^2}\theta } \right)}}{{{{\left( {{a^2} + 2{r^2} + {a^2}\cos 2\theta } \right)}^3}}}\\
\left( {\frac{{64\left( {{a^2} + {r^2} - 2f(r)} \right)f{{(r)}^2}{{\sin }^2}2\theta }}{{{{\left( {{a^2} + 2{r^2} + 
{a^2}\cos 2\theta  - 4f(r)} \right)}^4}}} + \frac{{{{\sin }^4}\theta {{\left( { - 4rf(r) + \left( {{a^2} + 2{r^2} + 
{a^2}\cos 2\theta } \right)f'(r)} \right)}^2}}}{{{{\left( {{a^2} + {r^2} - 2f(r) - {a^2}{{\sin }^2}\theta } \right)}^4}}}} \right).
\end{array}
\end{equation}
Now the active mass of the Kerr spacetime for the interior region is given by,
\begin{equation}
{{\cal M}_{GEM}}(r < R) = \int {\rho _{GEM}} d{\rm v} =  \int\limits_0^R {\int\limits_0^\pi  {\int\limits_0^{2\pi } {{\rho _{GEM}}\sqrt \gamma  drd\theta d\phi } } } 
\end{equation}
As was noticed, to calculate the above integral, we still need one more ingredient and that is the exact form of the $f(r)$ function. 
There are different choices for this function satisfying the boundary condition \eqref{BC}. To this end, we choose $f(r)$ such
that in the $a = 0$ limit, the $g_{00}$ of the interior Kerr metric reduces to that  of the interior Schwarzschild metric \cite{Col}.
This choice yields 
\begin{equation}
f(r) = \frac{{{r^2}}}{4}\left[ {\frac{{M{r^2}}}{{{R}^3}} + \frac{{9M}}{{{R}}} + 
3{{\left( {1 - \frac{{2M}}{{{R}}}} \right)}^{1/2}}{{\left( {1 - \frac{{2M{r^2}}}{{{R}^3}}} \right)}^{1/2}} - 3} \right]
\end{equation}
The integration over the coordinates covering the interior of the spheroidal leads to the following active mass for the interior region \footnote{There is a subtle point in taking the radial integral 
at the lower limit $r=0$ which gives zero. This is due to the fact that  unlike the spheroidal hypersurfaces $r=constant\neq0$, $r=0$ is a ring of radius $a$, $x^2 + y^2 = a^2$,  in the equatorial
plane ($z=0$).}
\begin{equation}
{{\cal M}_{GEM}}(r < R) = M + \frac{M^2}{R} + \frac{3}{2}\frac{{{M^3}}}{R^2} + 2\frac{{{M^2a^2}}}{R^3}  - \frac{91}{15} \frac{M^3a^2}{R^4} + {\rm higher\; orders}.
\end{equation}
Obviously, to have the Kerr active mass, this should be added to the active mass contribution from the exterior solution. 
The Kerr metric in Boyer-Lindquist coordinates is given by the following line element,
\begin{equation}
d{s^2} = (1 - \frac{{2Mr}}{{{\rho ^2}}})d{t^2} + \frac{{4Mar{{\sin }^2}\theta }}{{{\rho ^2}}}dt d\phi  - 
\frac{{{\rho ^2}}}{\Delta }d{r^2} - {\rho ^2}d{\theta ^2} - ({r^2} + {a^2} + 
\frac{{2M{a^2}r{{\sin }^2}\theta }}{{{\rho ^2}}}){\sin ^2}\theta d{\phi ^2}
\end{equation}
where
\begin{equation}
{\rho ^2} = {r^2} + {a^2}{{\cos }^2}\theta \;\;\;, \;\;\; \Delta  = {r^2} - 2Mr + {a^2}
\end{equation}
In the case of the exterior Kerr metric, we obtain the corresponding gravitoelectromagnetic fields,
\begin{equation}
\begin{array}{l}
{E_g}^2 = \frac{{{m^2}{{\rho ^2}}}}{{\left( \Delta \right){{\left( {2mr - {\rho ^2} } \right)}^2}}} + 
\frac{{4{a^4}{m^2}{r^2}{{\cos }^2}\theta {{\sin }^2}\theta }}{{{\rho ^2}{{\left( {{\rho ^2} - 2mr} \right)}^2}}}
\end{array}
\end{equation}
and
\begin{equation}
B_{g}^{2}=
\frac{{{a^2}{m^2}\left( {\Delta} \right)\left( {\rho ^2} - 2mr \right)\left( {\left( {{a^2} - 2{r^2} + 
{a^2}{{\cos }^2}2\theta } \right){{\sin }^4}\theta  + 4\Delta{r^2}{{\sin }^2}2\theta} \right)}}{{4{{\left( {{\rho ^2} - 2mr  } 
\right)}^4}\left( {4{m^2}{a^2}{r^2}{{\sin }^4}\theta  
+ \left( { {\rho ^2}- 2mr } \right){{\sin }^2}\theta \left( {\rho ^2}{\left( {{a^2} + {r^2}} \right)} + 2m{a^2}r{{\sin }^2}\theta  \right)} \right)}}
\end{equation}
which upon substitution in the exterior active mass density ${\rho _{GEM}}(r > {R}) = - \frac{1}{{4\pi }}{u_g}$  and computing the following integral
\begin{equation}
{{\cal M}_{GEM}}\left( {r > R} \right) =  - \frac{1}{{4\pi }}\int {(\frac{1}{2}{g_{00}}{B_g}^2 + {E_g}^2) } 
d{\rm v} =  - \frac{1}{{4\pi }}\int\limits_0^{2\pi } {\int\limits_0^\pi  {\int\limits_R^\infty {\left( {\frac{1}{2}{g_{00}}{B_g}^2 + {E_g}^2} \right)} } } 
\sqrt \gamma  drd\theta d\phi 
\end{equation}
and after expansions with respect to $\frac{a}{R}$ and $\frac{M}{R}$, we end up with
\begin{equation}
{{\cal M}_{GEM}}(r > R) = -\frac{{{M^2}}}{R} - \frac{3}{2}\frac{M^3}{R^2} - \frac{M^3a^2}{4R^4} + {\rm higher \; orders}.
\end{equation}
Therefore the (total) active mass of the kerr metric sums up to be
\begin{equation}
\begin{array}{l}
{{\cal M}_{GEM}} = {{\cal M}_{GEM}}(r < R) + {\cal M}_{GEM}(r > R) = 
M  + 2 \frac{{{M^2 a^2}}}{R^3} - \frac{379}{60} \frac{M^3 a^2}{R^4} +  {\rm higher \; orders \; terms}, 
\end{array}
\end{equation}
which in the limit $ a=0 $  is equal to the Schwarzschild active mass. At first sight, this may not be expected because the chosen interior Kerr 
metric  does not reduce to the interior Schwarzschild metric in the limit when the rotation parameter is set to zero. But one should recall that we also chose $f(r)$ such
that the $g_{00}$ of the interior Kerr metric reduces to that of the interior Schwarzschild metric for $a = 0$.


\section{Conclusion}
In the present study, after a brief introduction to the $1+3$ (threading) formulation of spacetime decomposition, the analogs of Poisson's equation in Einstein's  GR 
and its modification, i.e.; in the presence of a positive 
cosmological constant, are considered. Two different formalisms, one based on gravitoelectromagnetism and the resultant quasi-Maxwell form of the EFEs and the other 
one introduced by Ehlers et al. in \cite{Eh1}, are applied to different static and stationary spacetimes and the corresponding 
potentials and Poisson's equations are compared. Calculating the {\it active mass} (density) in both formalisms, it is shown that while in the case of 
GEM formalism one can identify the so-called gravitational binding/dispersing energy, the EOS  formalism does not accommodate such a concept. 
In the case of the Schwarzschild metric, it is found that in the GEM formalism the gravitational 
binding energy is larger than its value in the Newtonian approximation of the interior Schwarzschild solution. It is also shown that the inclusion of the cosmological constant leads to a positive 
binding energy (called gravitational dispersing energy) as expected from the repulsive  nature of gravity associated with the cosmological constant.\\
Noting that the EOS definition is confined to static spacetimes, we extend the calculation of GEM active density to the case of stationary spacetimes. As a prototype example, we
study the Kerr solution with a special choice for an interior solution 
matching the Kerr solution on a spheroidal hypersurface. It is found that the Kerr active mass finds contributions proportional to the rotation parameter, but reduces to the active 
mass of Schwarzschild  metric for $a=0$ . This is a
direct consequence of the fact that the interior Kerr is chosen such that its $g_{00}$ component reduces to that of the interior Schwarzschild solution in the same limit.\\
It should be noted that (local) mass is not a well-defined quantity in GR, and as such, there are different definitions of mass in the literature \cite{mass}. 
Alternative definitions of the so-called quasilocal mass, among many, include  Komar \cite {Kom}, ADM \cite {ADM}, Bondi-Sachs \cite {BS}, 
Penrose \cite {Pen}, Brown-York \cite {BY}, and Hawking-Horowitz \cite {HH} masses. 
These definitions, although they may or may not coincide for different spacetimes according to the spacetime properties, all share the same fact that they
are  defined {\it geometrically} in terms of integrals over closed 2-surfaces at spatial or null infinity \cite{Poiss}.
For example one can define the effective mass of the Kerr metric through a modification of the Komar mass as follows \cite{Kul}
 \begin{equation}\label{dad};
{M_{eff}} = M - \frac{{4{M^2}{a^2}}}{{{r^3}}} + \frac{{{M^3}{a^2}}}{{6{r^4}}} + ...
 \end{equation}
Another example is the quasilocal mass of the Kerr spacetime in the context of Brown-York quasilocal energy
\cite {BY}. In the slow rotation limit of the Kerr black hole for a 
constant radius surface $r=r_0$ with $r_0 \gg M$, it is given by \cite{Martinez}
\begin{equation}\label{mart}
E = M + \frac{M^2}{2r_0} + \frac{M^3}{2{r_0}^2} + \frac{5M^4}{8{r_0}^3} - \frac{7M^2 a^2}{6{r_0}^3} + {\cal O} (\frac{a^4}{{r_0}^3})
\end{equation}
Obviously unlike the active mass calculation which requires the interior Kerr solution, both of the 
above calculations are made using only the exterior Kerr metric. \\
To lift any ambiguities, it should be noted that the above definitions of quasilocal mass are based in one way or another on the $3+1$ 
(or slicing) decomposition formalism and the corresponding 
Hamiltonian formulation of GR so that the asymptotic two-dimensional (spatial or null) structure of the 
spacetime  determines the mass of the spacetime. On the other hand, in the $1+3$ (or threading) decomposition formalism and in analogy with Poisson's equation, it is a 3-volume 
integral over the active density which determines the active mass in {\it stationary} spacetimes. \\
As we pointed out, since in a curved background we do not have a consistent local
covariant definition of  energy (or mass) to which the property of conservation or nonconservation could  be assigned, people have 
tried to construct the so-called energy-momentum pseudotensor for the
gravitational field to be able to  account for the conservation of matter-energy including the gravitational field. In this regard, it would be
interesting to look for the relation between the GEM energy density discussed here and the components of proposed pseudotensors such as the Landau-Lifshitz 
energy-momentum pseudotensor \cite{Lan}. Finally, now that we have a  general relativistic analog of Poisson's equation for stationary spacetimes, using 
the methodology employed in \cite{Eh2}, one could look for the effect of the pressure term in stationary spacetimes and in particular whether or not there exists
any specific relation between the pressure changes inside the source and stress changes on its boundary. Obviously, this would be very interesting to consider in the case of 
a rotating body represented by an interior Kerr metric and the corresponding equation of state.
\section *{Acknowledgments}
The authors would like to thank the University of Tehran for supporting this project under the grants provided by the research council. M. N.-Z also thanks the high energy, cosmology 
, and astroparticle group at the Abdus Salam ICTP for kind hospitality during his visit when part of this study was carried out.




\begin{thebibliography}{99} 

\bibitem{Eh1} J. Ehlers, I. Ozsvath and  E. Schucking,  Am. J. Phys. {\bf74}, 607 (2006).

\bibitem{Steph} H. Stephani, {\it Introduction to General Relativity} (Cambridge University Press, Cambridge, England, 1992).

\bibitem{Eh2} J. Ehlers et al., Phys. Rev. D {\bf 72}, 124003 (2005).

\bibitem{Wald} R. M. Wald,  {\it General Relativity} (University of Chicago Press, Chicago, 1984) pp. 126-127.

\bibitem{ryder} L. Ryder,  {\it Introduction to General Relativity } Cambridge University Press, Cambridge, England, 2009.

\bibitem{Lan} L. D. Landau and  E. M. Lifshitz, {\it The Classical Theory of Fields} (Pergamon, New York, 1975).

\bibitem{Tolman} R. C. Tolman, {\it Relativity, Thermodynamics and Cosmology} (Clarendon Press, Oxford, 1934).

\bibitem{Misner} C. W. Misner and P. Putnam, Phys. Rev. {\bf 116}, 1045 (1959).

\bibitem{MN} D. Lynden-Bell and M. Nouri-Zonoz, Rev. Mod. Phys. {\bf 70}, 427 (1998).

\bibitem{Nouri99} M. Nouri-Zonoz, Ph.D. thesis, University of Cambridge, 1998.

\bibitem{Nouriz}
M. Nouri-Zonoz, Phys. Rev. D {\bf 60}, 024013 (1999);
M. Nouri-Zonoz and B. Nazari, Phys. Rev. D {\bf 82}, 044047 (2010);
M. Nouri-Zonoz and B. Nazari, Phys. Rev. D {\bf 85}, 044060 (2012); 
M. Nouri-Zonoz and A. Parvizi, Phys. Rev. D {\bf 88}, 023004 (2013).

\bibitem{Fil} L. Filipe, O. Costa, and J. Natario, Gen. Relativ. Gravit. {\bf 46}, 1792 (2014).

\bibitem{NKR} M. Nouri-Zonoz, J. Koohbor, and H. Ramezani-Aval, Phys. Rev. D, {\bf 91}, 063010 (2015).

\bibitem{stu} Z. Stuchlik, Acta Phys. Slovaca {\bf 50}, 219-228, (2000).

\bibitem{Gi} M. G\"{u}rses and  F. G\"{u}rsey, J. Math. Phys. {\bf 16}, 2385 (1975).


\bibitem{Col} P. Collas and J. K. Lawrence, Gen. Relativ. Gravit. {\bf 7}, 715 (1976).

\bibitem{mass} L. B. Szabados, Living Rev. Relativity {\bf 12}, {\bf 4} (2009).

\bibitem{Kom}  A. Komar, Phys. Rev. {\bf 113}, 934 (1959).

\bibitem{ADM}  R. Arnowitt, S. Deser, and C.W. Misner, in {\it Gravitation: An introduction to current research}, edited by 
L. Witten (Wiley, New York, 1962), pp. 227-265.

\bibitem{BS} H. Bondi, Nature (London) {\bf 186}, 535 (1960); H. Bondi, M. G. J. van der Burg and A. W. K. Metzner, 
Proc. R. Soc. A. {\bf 269}, 21 (1962); R. K. Sachs, Proc. R. Soc. A {\bf 270}, 103 (1962).

\bibitem{Pen} R. Penrose, Proc. R. Soc. Lond. A {\bf 381}, 53, 1982. 

\bibitem{BY} J. D. Brown and  J. W. York, Phys. Rev. {\bf D} 47, 1407-1419 (1993).

\bibitem{HH} S. W. Hawking and G. T. Horowitz, Classical and Quantum Gravity 13, 1487-1498 (1996).

\bibitem{Poiss} E. Poisson, {\it A relativist's toolkit}, Cambridge University Press, Cambridge, UK, 2004. 

\bibitem{Kul} R. Kulkarni, V. Chellathurai and N. K. Dadhich, Classical and Quantum Gravity, 5(11), p.1443 (1988).

\bibitem{Martinez} E. A. Martinez, Phys. Rev. {\bf D} 50, 4920 (1994).


\end{thebibliography}
\end{document}